\documentclass[pra,a4paper,preprint,showpacs,superscriptaddress]{revtex4}
\usepackage{amsmath}
\usepackage{amsfonts}
\usepackage{graphicx}
\usepackage{longtable}
\newcommand{\be}{\begin{equation}}
\newcommand{\ee}{\end{equation}}

\newcommand{\ba}[1]{\left(\begin{array}{#1}}
\newcommand{\ea}{\end{array}\right)}
\begin{document}

\title{Monogamy of quantum correlations in three qubit pure states using Rajagopal-Rendell (RR) quantum deficit} 
\author{Sudha}
\email{arss@rediffmail.com}
\affiliation{Department of Physics, Kuvempu University, Shankaraghatta, Shimoga-577 451, India.}
\affiliation{Inspire Institute Inc., Alexandria, Virginia, 22303, USA.}
\author{A.R.Usha Devi}
\affiliation{Inspire Institute Inc., Alexandria, Virginia, 22303, USA.}
\affiliation{Department of Physics, Bangalore University, 
Bangalore-560 056, India}
\author{A.K.Rajagopal} 
\affiliation{Inspire Institute Inc., Alexandria, Virginia, 22303, USA.}
 \date{\today}
\begin{abstract} 
The limitation on the shareability of quantum entanglement over several parties, the so-called monogamy of entanglement, is an issue that has caught considerable attention of quantum information community over the last decade. A natural question of interest in this connection is whether monogamy of correlations is true for correlations other than entanglement. This issue is examined here by choosing quantum deficit, proposed by Rajagopal and Rendell, an operational measure of correlations. In addition to establishing the polygamous nature of the class of three qubit symmetric pure states characterized by two distinct Majorana spinors (to which the W states belong), those with three distinct Majorana spinors (to which GHZ states belong) are shown to either obey or violate monogamy relations.  While the generalized W states can be mono/polygamous, the generalized GHZ states exhibit monogamy with respect to quantum deficit. The issue of using monogamy conditions based on quantum deficit to witness the states belonging to SLOCC inequivalent classes is discussed in the light of these results.  
\end{abstract}
\maketitle

\section{Introduction}
One among the several features that distinguish correlations in quantum and classical scenario is their shareability amongst several parties of a composite system. While classical correlations are known to be infinitely shareable, quantum correlations, especially quantum entanglement has limited shareability. Starting from the seminal work of Coffman, Kundu and Wootters~\cite{ckw}, many researchers working in quantum information theory are addressing this issue, the so called monogamy of entanglement~\cite{ter,mixed,gauss,renyl}. 

Quite recently, enquiries regarding how non-classical correlations~\cite{oz,akr,qc}, other than entanglement -- characterized by the measure quantum discord~\cite{oz} -- get shared amongst more than two parties  have been raised~\cite{prabhu,gl}. From these works it is found that  the quantum correlations in three qubit pure states do not necessarily obey  any stringent monogamy inequality (as inferred through  quantum discord)~\cite{prabhu,gl}. 

In the present work, quantum deficit~\cite{akr} proposed by Rajagopal and Rendell (henceforth called RRQD) is chosen as the measure of quantum correlations. This measure is particularly chosen because, unlike quantum discord, no optimization is required to evaluate RRQD, thus making it an operationally convenient measure of quantum correlations~\cite{akr,qc}. The monogamy of RRQD for the two SLOCC inequivalent classes of $3$ qubit symmetric pure states characterized by two and three distinct spinors respectively is examined. The generalized class of GHZ and W states are also explored for their mono/polygamous nature. While it can be conclusively shown that any state belonging to the two-distinct spinor class is
polygamous, states belonging to the 3-distinct spinor class can either be monogamous or
polygamous. The generalized class of GHZ states can be shown to obey the monogamy relations, quite like the GHZ states themselves but the states belonging to generalized W class can either obey or violate them. 

The paper is organized as follows: In Sec II, a brief introduction to the concept of  RRQD  is given and the classification of symmetric pure states based on the distinct spinors characterizing them is detailed in Sec III.  The monogamy relation with respect to RRQD is given in Sec.IV and it is shown that all states belonging to the $2$-distinct spinor class violate this relation.  Two examples of pure symmetric states characterized by $3$ distinct spinors, one of them being the GHZ state, are considered and it is shown that GHZ state is monogamous whereas the other state(belonging to the same SLOCC class) is not. The generalized class of GHZ and W states are examined for their mono/polygamous nature in Sec.V.  Sec. VI contains a summary and discussion of the results.  

\section{RR Quantum Deficit} 
In their enquiry into the circumstances under which entropy methods can give an answer to the
questions of quantum separability and classical correlations of a composite state, Rajagopal and Rendell~\cite{akr} proposed a useful measure of quantum correlations, the {\em{quantum deficit}} (which is referred to as RRQD to distinguish it from one-way quantum deficit~\cite{jo}, another measure of quantum correlations). The RRQD of a bipartite quantum state $\rho_{AB}$ is defined~\cite{akr} as the relative entropy~\cite{nc} of the state $\rho_{AB}$ with its classically decohered counterpart $\rho^d_{AB}$ as follows: 
\be
D_{AB}=S(\rho_{AB}\vert\vert \rho^d_{AB})
\ee
where 
\be
S(\rho_{AB}\vert\vert \rho^d_{AB})=\mbox{Tr} (\rho_{AB} \log \rho_{AB})-\mbox{Tr} (\rho_{AB} \log \rho^d_{AB}). 
\ee
Quantum deficit $D(A,B)$ determines the quantum excess of correlations in the state $\rho_{AB}$ with reference to its classically decohered counterpart $\rho_{AB}^d$. The state $\rho^d_{AB}$  shares the same subsystems $\rho_A$, $\rho_B$ as that of $\rho_{AB}$ and is diagonal in the eigenbasis $\{\vert a\rangle\}$, $\{\vert b\rangle\}$ of $\rho_A$, $\rho_B$:  
\be 
\rho^d_{AB}=\sum_{a,\,b} P_{ab}\, \vert a\rangle \langle a\vert \otimes \vert b\rangle \langle b\vert=\sum_{a,\,b} P_{ab}\, \vert a,\,b \rangle   \langle a,\, b\vert
\ee
where $P_{ab}=\langle a, b\vert\rho_{AB}\vert a, b\rangle$ denote the diagonal elements of $\rho_{AB}$ and $\sum_{a,b}\ P_{ab}=1$.

It may be readily seen that 
\begin{eqnarray}
\mbox{Tr} (\rho_{AB} \log \rho^d_{AB})&=&\mbox{Tr} (\rho^d_{AB} \log \rho^d_{AB})\nonumber \\ 
&=& \sum_{a,b} P_{ab}\log P_{ab}. 
\end{eqnarray} 
We thus obtain 
\begin{eqnarray}
\label{newd}
D(A,B)&=&\mbox{Tr} (\rho_{AB} \log \rho_{AB})-\mbox{Tr} (\rho_{AB} \log \rho^d_{AB})\nonumber \\ 
&=& \sum_i \lambda_i \log \lambda_i-\sum_{a,b} P_{ab}\log P_{ab}, 
\end{eqnarray}
where $\lambda_i$ denote the eigenvalues of the bipartite state $\rho_{AB}$. 

It is pertinent to point out that quantum discord~\cite{oz} and one way quantum deficit~\cite{jo} are asymmetric with respect to measurements on the subsystem $A$ and $B$ whereas, RRQD is symmetric about the subsystems $A$, $B$. Another advantage of RRQD over the other two measures is that it does not require any optimal measurement schemes for its evaluation. An explicit evaluation of RRQD for some classes of $3$ qubit pure states  is carried out in this paper and the shareability aspects of quantum correlations among its $2$ qubit subsystems are examined.      

\section{Classification of pure symmetric states based on Majorana representation} 
An elegant representation of $N$ qubit pure symmetric states is given by Majorana~\cite{maj} way back in 1932. In addition to the possibility of expressing symmetric pure states as $N$ points on the block sphere, there are several advantages of resorting to this representation. 
Here is a brief introduction to the Majorana representation of $N$ qubit pure symmetric states, in order to facilitate its use in this paper.  

Majorana~\cite{maj} proposed that a pure symmetric state of  $N$ spinors (a pure state of spin $j=\frac{N}{2}$)  quantum state can essentially be represented as a {\em symmetrized} combination of $N$ constituent spinors as  
\begin{equation}
\label{Maj}
\vert \Psi_{\rm sym}\rangle={\cal N}\, \sum_{P}\, \hat{P}\, \{\vert \epsilon_1, \epsilon_2, 
\ldots  \epsilon_N \rangle\}, 
\end{equation} 
where 
\begin{equation}
\label{spinor}
\vert\epsilon_l\rangle= 
\cos(\beta_l/2)\, e^{-i\alpha_l/2}\, \vert 0\rangle +
\sin(\beta_l/2)\, e^{i\alpha_l/2} \, \vert 1\rangle,\ \ l=0,1,2,\ldots, N,
\end{equation}
denote the spinors constituting the state $\vert \Psi_{\rm sym}\rangle$;
$\hat{P}$ corresponds to the set of all $N!$ permutations of the spinors (qubits) and ${\cal N}$ corresponds to an overall normalization factor. 

Bastin et.al~\cite{bastin} made use of the representation~(\ref{Maj}) for the classification of pure symmetric $N$ qubit states into SLOCC inequivalent classes. The classification is based on the number of distinct spinors (degeneracy number) and their frequency of occurence (degeneracy configuration) in the state under consideration. A comprehensive review of this classification and the uses of Majorana representation
may be found in Ref.~\cite{usrmaj}. 

An $N$ qubit state containing $r(<N)$ distinct spinors $\vert \epsilon_i \rangle$ ($i=1,\,2,\,\ldots\, r$), each repeating $n_i$ times,  belongs to the class ${\cal D}_{n_1,\,n_2,\,\ldots\,n_r}$ and each degeneracy configuration $\{n_1,\,n_2,\,\ldots\,n_r \}$ (with the numbers $n_i$ being arranged in the descending order) corresponds to a {\em distinct} SLOCC class. The number of SLOCC inequivalent classes possible for states with $r$ distinct spinors is given by the partition function $p(N,\,r)$ that gives the distinct possible ways in which the number  $N$ can be partitioned into $r$ numbers $n_i$  ($i=1,\,2,\,\ldots\, r$) such that $\sum_{i=1}^r \, n_i=N$. For instance, a $3$ qubit state with only one distinct spinor belongs to the class ${\cal D}_3$,  with two distinct spinors belongs to the class $\{{\cal D}_{2,1}\}$ and $\{{\cal D}_{1,1,1}\}$ is the class of $3$ qubit states with three distinct spinors. The classes ${\cal D}_{3}$, ${\cal D}_{2,1}$ and ${\cal D}_{1,1,1}$ are SLOCC inequivalent and a state belonging to one of these classes cannot be converted into the other (different from itself) by any local operations and classical communications. 

A representative  symmetric state with two distinct spinors belonging to the entanglement family 
$\{ {\cal D}_{N-k,k}, k=1,2,\ldots, [N/2]\}$ is given by 
\begin{eqnarray}
\label{dnk}
\vert \Psi_{N-k, k}\rangle &=& {\cal N}\, \sum_{P}\, \hat{P}\,\{ \vert \underbrace{\epsilon_1, \epsilon_1,
\ldots , \epsilon_1}_{N-k};\ \underbrace{\epsilon_2, \epsilon_2,\ldots , \epsilon_2}_{k}\rangle\}\nonumber \\
&=& {\cal N}\, R_1^{\otimes N}\, \sum_{P}\, \hat{P}\,\{ \vert \underbrace{0, 0,
\ldots , 0}_{N-k};\ \underbrace{\epsilon'_2, \epsilon'_2,\ldots , \epsilon'_2}_{k}\rangle\},
\end{eqnarray}
where $\epsilon_1=R_1\vert 0\rangle$ and $\epsilon_2=R_2\vert 0\rangle$, and 
\begin{equation}
\label{ep'}
\vert \epsilon'_2\rangle=R_1^{-1}R_2\vert 0\rangle=d_0\, \vert 0\rangle+d_1\, \vert 1\rangle,\ \ \vert d_0\vert^2+\vert d_1\vert^2=1,\ \ d_1\neq0.
\end{equation}
Substitution of (\ref{ep'}) into (\ref{dnk}) and further simplication leads to 
\be
\label{3q}
\vert \Psi_{N-k, k}\rangle = R_1^{\otimes N}\, \sum_{r=0}^k\,\sqrt{^N C_{r}}\, \alpha_{r}\, \left\vert\frac{N}{2},\frac{N}{2}-r \right\rangle,
\ \ {\rm where} \  \ \alpha_{r}={\cal N}\,\, 
\frac{(N-r)!}{(N-k)! (k-r)!}\, d_0^{k-r}\, d_1^r. 
\ee
This implies that all symmetric states $\vert \Psi_{N-k, k}\rangle$, constituted by two distinct Majorana spinors, are equivalent 
(under local unitary transformations) to 
\begin{equation}
\label{nk'}
\vert \Psi'_{N-k, k}\rangle=R_1^{-1\, \otimes N}\vert D_{N-k, k}\rangle=\sum_{r=0}^k\, \sqrt{^N C_{r}}\, \alpha_{r}\, \left\vert\frac{N}{2},\frac{N}{2}-r \right\rangle.%\equiv \vert \Psi_{N-k, k}\rangle.
\end{equation}
As $d_1\neq 0$, the coefficients  $\alpha_r$, ($r=0,\,1,\,2,\ldots,\,k$) are non-zero, except when $d_1=1, d_0=0$ -- in which case the state  $\vert D'_{N-k, k}\rangle$ reduces to the Dicke state $\left\vert\frac{N}{2},\frac{N}{2}-k \right\rangle$ itself and then, 
$\alpha_r=\delta_{k,r}$. 

An arbitrary $3$ qubit pure symmetric state $\vert \psi \rangle$ with two distinct spinors is given by (obtained by substituting $N=3$ and $k=1$ in   (\ref{nk'})),  
\begin{eqnarray}
\label{3qu} 
\vert \psi \rangle&=&\sum_{r=0}^1\, \sqrt{^3 C_{r}}\, \alpha_{r}\, \left\vert\frac{3}{2},\frac{3}{2}-r \right\rangle =\alpha_0 \left\vert\frac{3}{2},\frac{3}{2} \right\rangle+\sqrt{3} \alpha_1 \left\vert\frac{3}{2},\frac{1}{2} \right\rangle.
\end{eqnarray}
which may be expressed in terms of standard qubit basis as, 
\be
\label{3q2}
\vert \psi \rangle \equiv a \vert 000 \rangle+ b \left(\frac{\vert 100 \rangle+\vert 010 \rangle+\vert 001 \rangle}{\sqrt{3}}\right)    
\ee   
with $a=\alpha_0$, $b=\sqrt{3}\,\alpha_1$ are complex numbers obeying $\vert a \vert^2+\vert b \vert^2=1$. On taking $a=\cos \frac{\theta}{2}$, $b=\sin \frac{\theta}{2}\, e^{i\phi}$, ($0<\theta<\pi$, $0<\phi< 2\pi$),  without any loss of generality and subjecting the three qubit state (\ref{3q2})  to another local unitary transformation $\vert 0\rangle'=\vert 0\rangle,\ \ \vert 1\rangle'=e^{-i\phi}\vert 1\rangle$ on all the three qubits we obtain a further simplified form  
\be
\label{sim}
\vert \psi' \rangle \equiv \cos \frac{\theta}{2} \vert 000 \rangle+ \sin \frac{\theta}{2} \left(\frac{\vert 100 \rangle+\vert 010 \rangle+\vert 001 \rangle}{\sqrt{3}}\right) 
\ee
with a single parameter $\theta$, $0<\theta<\pi$ describing the state.    
 
This general form (\ref{sim}) of a pure symmetric $3$ qubit state containing two distinct spinors will be made use of in the evaluation of RRQD for the state and its $2$ qubit subsystems in the next section.  

\section{Monogamy of $3$ qubit pure symmetric states with respect to RRQD} 

It is well known that~\cite{ckw,ter} monogamy relations capture the trade-off between the quantum correlations in the subsystems of a composite state and that in the whole state. The monogamy inequality with respect to a measure, say $\cal Q$, of quantum correlations is given by  
\be
\label{monoq}
{\cal Q}(\rho_{A:B})+{\cal Q}(\rho_{A:C})\leq {\cal Q}(\rho_{A:BC})
\ee
for a tripartite state $\rho_{ABC}$. Here ${\cal Q}(\rho_{A:B})$, ${\cal Q}(\rho_{A:C})$ denote the correlations in the states $\rho_{AB}=\mbox{Tr}_{C}\, \rho_{ABC}$, $\rho_{AC}=\mbox{Tr}_{B}\,\rho_{ABC}$ respectively and  ${\cal Q}(\rho_{A:BC})$ gives the tripartite correlation, between the subsystem $A$, $BC$ of $\rho_{ABC}$. 
 
The measure of correlations adopted here being the RRQD, the monogamy inequality (\ref{monoq}) may be expressed as, 
\be
\label{monoqrr}
D_{AB}+D_{AC}\leq D_{A:BC}
\ee 
A tripartite state  $\rho_{ABC}$ is monogamous with respect to RRQD iff the inequality (\ref{monoqrr}) is obeyed and polygamous  otherwise. 

\subsection{$3$-qubit pure symmetric states with $2$ distinct spinors} 

Considering a pure symmetric state $\vert \psi \rangle$ of $3$ qubits constituted by $2$ distinct spinors, having the structure given in Eq.(\ref{sim}), it is not difficult to see that $\rho_{AB}=\rho_{BC}$, owing to the symmetry of the state. Thus, $D_{AB}=D_{AC}$. 
An explicit evaluation of eigenvectors of the subsystems $\rho_A(= \rho_B$) of $\rho_{AB}=\mbox{Tr}_{C}\,\vert \psi \rangle \langle \psi \vert$ leads to $\rho_{AB}^d$, the completely decohered counterpart of $\rho_{AB}$. In fact the diagonal entries of $\rho_{AB}^d$ are given by  
\begin{eqnarray}
P_{11}&=&\langle \chi_1,\chi_1\vert \rho_{AB} \vert \chi_1, \chi_1\rangle=\frac{1}{24}\left[14+\cos \theta+4\sqrt{6+4 \cos \theta-\cos 2\theta}-\frac{9(2+\cos \theta)}{6+4 \cos \theta-\cos 2\theta} \right] \nonumber \\
P_{12}&=&\langle \chi_1,\chi_2\vert \rho_{AB} \vert \chi_1, \chi_2\rangle=\frac{(2+\cos \theta) \sin^4 \frac{\theta}{2}}{3(6+4 \cos \theta-\cos 2\theta)}\nonumber \\
P_{21}&=&\langle \chi_2,\chi_1\vert \rho_{AB} \vert \chi_2, \chi_1\rangle=\frac{(2+\cos \theta) \sin^4 \frac{\theta}{2}}{3(6+4 \cos \theta-\cos 2\theta)} \nonumber \\
P_{22}&=&\langle \chi_2,\chi_2\vert \rho_{AB} \vert \chi_2, \chi_2\rangle=\frac{1}{24}\left[14+\cos \theta-4\sqrt{6+4 \cos \theta-\cos 2\theta}-\frac{9(2+\cos \theta)}{6+4 \cos \theta-\cos 2\theta} \right]. \nonumber 
\end{eqnarray}
Here $\vert \chi_i \rangle$, $i=1,\,2$ denote the eigenvectors of $\rho_A(= \rho_B$) belonging to the eigenvalues 
\be
\lambda_1=\frac{1}{6}(3+\sqrt{6+4 \cos \theta-\cos 2\theta}), \   \lambda_2=\frac{1}{6}(3-\sqrt{6+4 \cos \theta-\cos 2\theta})
\ee 
respectively. The relative entropy $S(\rho_{AB}\vert\vert\rho_{AB}^d)$ (See Eq.(\ref{newd})) can thus be arrived at and the RRQD $D_{AB}=S(\rho_{AB}\vert\vert\rho_{AB}^d)$ is obtained as a function of the parameter $\theta$. Similarly the eigenvectors of the subsystems $\rho_{BC}$, $\rho_A$ of $\rho_{ABC}=\vert \psi \rangle \langle \psi \vert$ facilitate the evaluation of $\rho_{ABC}^d$. 
Denoting the eigenvectors of $\rho_{BC}$ by $\vert \eta_i\rangle$, $i=1,\,2,\,3,\,4$, one has
$\left(\rho^d_{A:BC}\right)_{ij;ij}=\langle \chi_i,\eta_j\vert \rho_{ABC} \vert \chi_i, \eta_j\rangle$ and it turns out that 
\be
\rho^d_{A:BC}=\mbox{diag} \, \left(0,\,0,\,0,\,0,\,\frac{3+\sqrt{6+4 \cos \theta-\cos 2\theta}}{6},\,0,\,0,\,\frac{3-\sqrt{6+4 \cos \theta-\cos 2\theta}}{6}\right). \nonumber 
\ee 
Since $S(\rho_{ABC})$ is zero as $\rho_{ABC}$ corresponds to a pure state, the RRQD $D_{A:BC}=S(\rho_{ABC}\vert\vert\rho_{ABC}^d)$ is given by 
\begin{eqnarray}
D_{A:BC}&=&0-\frac{3+\sqrt{6+4 \cos \theta-\cos 2\theta}}{6}\log \frac{3+\sqrt{6+4 \cos \theta-\cos 2\theta}}{6} \nonumber \\
        && -\frac{3-\sqrt{6+4 \cos \theta-\cos 2\theta}}{6}\log \frac{3-\sqrt{6+4 \cos \theta-\cos 2\theta}}{6}. 
\end{eqnarray}
It may be readily seen that the state $\rho_{ABC}=\vert \psi \rangle \langle \psi \vert$ is monogamous iff 
\be
Q_{\rm II}=D_{AB}+D_{AC}-D_{ABC}=2 D_{AB}-D_{A:BC} \leq 0
\ee  
A plot of $Q_{\rm II}$ versus $\theta$ is shown in Fig.1

\begin{figure}[h]
\includegraphics*[width=2.2in,keepaspectratio]{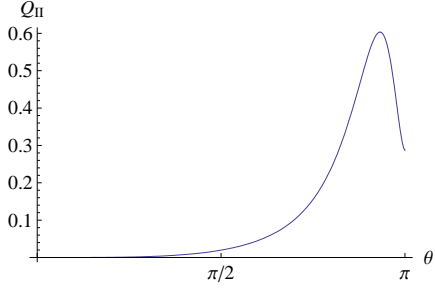}
\caption{A plot of $Q_{\rm II}=D_{AB}+D_{AC}-D_{ABC}=2 D_{AB}-D_{A:BC}$ versus $\theta$ for $3$ qubit pure states with two distinct Majorana spinors. The positive values of $Q_{\rm II}$ indicate violation of the monogamy relation.} 
\end{figure}   
It can be seen through the graph that $Q_{\rm II}>0$ for the whole range implying that {\emph{all the $3$ qubit pure states with two distinct majorana spinors do not obey the monogamy inequality and thus are polygamous}}.   

\subsection{$3$-qubit pure symmetric states with $3$ distinct spinors} 

Having established the polygamous nature of the set of all $3$-qubit pure states belonging to the SLOCC class ${\cal D}_{2,1}$, two specific examples of the $3$ qubit states belonging to the SLOCC class ${\cal D}_{1,1,1}$ with $3$ distinct spinors will now be considered. The states under consideration are 
\begin{eqnarray}
\label{ghz}
\vert{\rm GHZ} \rangle &=& \frac{\vert 000 \rangle+\vert 111 \rangle}{\sqrt{2}}  \\ 
\label{ww}
\vert \rm{ W\bar{W}} \rangle &=& \frac{\vert 100 \rangle+\vert 010 \rangle+\vert 001 \rangle+\vert 011 \rangle+\vert 101 \rangle+\vert 110  \rangle}{\sqrt{6}}=\frac{\vert W \rangle +\vert {\bar W} \rangle}{\sqrt{2}} 
\end{eqnarray}
with $\vert {\rm W} \rangle=\frac{\vert 100 \rangle+\vert 010 \rangle+\vert 001 \rangle}{\sqrt{3}}$, $\vert {\rm {\bar W}} \rangle=\frac{\vert 011 \rangle+\vert 101 \rangle+\vert 110 \rangle}{\sqrt{3}}$ being the $W$, obverse $W$ states. It can be seen that~\cite{usrmaj} the state $\vert{\rm GHZ} \rangle$ is comprised of the spinors 
\be
\vert \epsilon_1\rangle =\frac{1}{\sqrt{2}}[\vert 0\rangle+\omega\, \vert 1\rangle], \ \ 
\vert \epsilon_2\rangle =\frac{1}{\sqrt{2}}[\vert 0\rangle+\omega^2\, \vert 1\rangle], \ \ 
\vert \epsilon_3\rangle =\frac{1}{\sqrt{2}}[\vert 0\rangle+ \vert 1\rangle] 
\ee  
with $\omega,\  \omega^2,\  \omega^3=1$ being the cube-roots of unity 
and the spinors 
\be
\vert \epsilon'_1\rangle =\vert 1\rangle,\ \  \vert \epsilon'_2\rangle =\frac{\vert 0\rangle+ \vert 1\rangle}{\sqrt{2}} \ \  
\vert \epsilon'_3\rangle = \vert 0\rangle  
\ee
constitute the state $\vert {\rm{W{\bar W}}}\rangle$. Thus both $\vert {\rm GHZ }\rangle$, $\vert \rm{W{\bar W}}\rangle$ belong to the SLOCC class 
${\cal D}_{1,1,1}$ and a check whether both these states obey monogamy relation with respect to RRQD will now be carried out.  

It is known that $\vert {\rm GHZ} \rangle$ exhibits genuine 3-party entanglement (three tangle $\tau=1$ and vanishing pairwise concurrence), whereas  the state $\vert \rm{ W\bar{W}} \rangle$ possesses both 3-party entanglement (three tangle $\tau=1/3$) and also pairwise entanglement (with concurrence of two qubit marginal states $C=1/3$)~\cite{usa2}. 

As there are no two-qubit correlations in $\vert {\rm GHZ} \rangle$, $D_{AB}=D_{AC}=0$. The completely decohered counterpart $\rho^d_{\rm GHZ}$ of the GHZ state is diagonal in the eigenbasis of subsystems $A$ and $BC$  with $\frac{1}{2}$, $\frac{1}{2}$ as the non-zero diagonal elements and one gets $D(A:BC)=\log 2$. Thus, $\vert {\rm GHZ} \rangle$ obviously satisfies the monogamy relation (\ref{monoqrr}) establishing its monogamous nature. 

Considering the state $\vert {\rm{W{\bar W}}}\rangle$, it is readily seen that 
\be
\rho_{AB}=\rho_{AC}=\frac{1}{6}\ba{cccc} 1 & 1 & 1 & 0 \\ 1 & 2 & 2 & 1 \\ 1 & 2 & 2 & 1 \\0 & 1 & 1 & 1 \ea
\ee
and 
\begin{eqnarray}
\rho_A&=&\mbox{Tr}_B  {\rho_{AB}}=\mbox{Tr}_C {\rho_{AC}}=\frac{1}{6}\ba{cc} 3 & 2 \\ 2 & 3 \ea=\rho_B=\rho_C
\end{eqnarray}
in the standard qubit basis $\{\vert 00\rangle, \vert 01\rangle,\ \vert 10\rangle, \vert 11\rangle\}$. 
 
The eigenvectors of  $\rho_A=\mbox{Tr}_{BC} {\rho_{\rm W{\bar W}}}$ $=(\rho_B=\mbox{Tr}_{AC} {\rho_{\rm W{\bar W}}})$ are given by, 
\be
\vert \chi_1\rangle=\frac{1}{\sqrt{2}}[\vert 0\rangle+\vert 1\rangle)\in \lambda_1=\frac{5}{6},\ \  \vert \chi_2\rangle=\frac{1}{\sqrt{2}}[\vert 0\rangle-\vert 1\rangle)\in \lambda_2=\frac{1}{6}
\ee   
and the decohered state $\rho^d_{AB}$ can be identified as,  
\begin{scriptsize}
\begin{eqnarray}
\rho^d_{AB}&=&\rho^d_{AC} \nonumber \\
&=&\mbox{diag}\,\left( \langle \chi_1,\chi_1\vert \rho_{AB} \vert \chi_1, \chi_1\rangle, \, \langle \chi_1,\chi_2\vert \rho_{AB} \vert \chi_1, \chi_2\rangle\, \langle \chi_2,\chi_1\vert \rho_{AB} \vert \chi_2, \chi_1\rangle, \, \langle \chi_1,\chi_2\vert \rho_{AB} \vert \chi_2, \chi_2\rangle                       \right) \nonumber \\ 
&=&\mbox{diag}\,\left(\frac{3}{4},\,\frac{1}{12},\,\frac{1}{12},\,\frac{1}{12}\right)   
\end{eqnarray}
\end{scriptsize} 
The non-zero eigenvalues of $\rho_{AB}$ here being $\lambda_1=\frac{5}{6}$, $\lambda_2=\frac{1}{6}$, we obtain RRQD to be,  
\begin{eqnarray}
D_{AB}&=&\frac{5}{6}\log \frac{5}{6}+\frac{1}{6}\log \frac{1}{6} -\left(\frac{3}{4} \log \frac{3}{4}+\frac{3}{12}\log \frac{1}{12}\right) \approx 0.386 
\end{eqnarray}

In order to evaluate $D_{A:BC}$, we proceed to evaluate the completely decohered counterpart $\rho^d_{A:BC}$ of the state $\rho_{A:BC}=\rho_{W\bar{W}}$ under consideration:  The eigenvectors of bipartite subsystems $\rho_{BC}=\rho_{AB}$ of $\rho_{\rm W{\bar W}}$ are given by,  
\begin{eqnarray*}
\vert \eta_1\rangle&=&\frac{1}{\sqrt{10}}[\vert 0,0\rangle +2\, \vert 0,1\rangle+2\, \vert 1,0\rangle + \vert 1,1\rangle], \\ \vert \eta_2\rangle&=&\frac{1}{\sqrt{2}}[\vert 0,0\rangle  - \vert 1,1\rangle],  \\
\vert \eta_3\rangle&=&\frac{1}{\sqrt{3}}[\vert 0,0\rangle -\, \vert 0,1\rangle + \vert 1,1\rangle], \\ 
\vert \eta_4\rangle&=&\frac{1}{\sqrt{2}}[ \vert 0,1\rangle - \vert 1,0\rangle].
\end{eqnarray*}
The decohered counterpart $\rho^d_{A:BC}$ of the three qubit state $\rho_{W\bar{W}}$ is given by, 
\begin{eqnarray}
\rho^d_{A:BC}&=&\mbox{diag}\,\left( \langle \chi_1,\eta_1\vert \rho_{\rm W{\bar W}} \vert \chi_1, \eta_1\rangle, \langle \chi_1,\eta_2\vert \rho_{\rm W{\bar W}} \vert \chi_1, \eta_2\rangle, \right. \nonumber \\
&& \ \ \langle \chi_1,\eta_3\vert \rho_{\rm W{\bar W}} \vert \chi_1, \eta_3\rangle, \langle \chi_1,\eta_4\vert \rho_{\rm W{\bar W}} \vert \chi_1, \eta_4\rangle, \nonumber \\ 
&& \ \langle \chi_2,\eta_1\vert \rho_{\rm W{\bar W}} \vert \chi_2, \eta_1\rangle, \langle \chi_2,\eta_2\vert \rho_{\rm W{\bar W}} \vert \chi_2, \eta_2\rangle,  \nonumber \\
&&\left.  \ \ \langle \chi_2,\eta_3\vert \rho_{\rm W{\bar W}} \vert \chi_2, \eta_3\rangle, \langle \chi_2,\eta_4\vert \rho_{\rm W{\bar W}} \vert \chi_2, \eta_4\rangle\right)\nonumber \\ 
&=& \mbox{diag} \, (\frac{5}{6},\,0,\,0,\,0,\,0,\,(\frac{1}{6},\,0,\,0)
\end{eqnarray} 
The  RRQD $D(A:BC)$ characterizing the quantum correlations between $A$ and $BC$ subsystems of the state  $\rho_{W\bar{W}}$ is then obtained as,
\be
D_{A:BC}=0-\left(\frac{5}{6}\log \frac{5}{6}+\frac{1}{6}\log \frac{1}{6} \right) \approx 0.45.
\ee
Finally, it may be seen that the monogamy inequality (\ref{monoqrr})  governing the pairwise and three party correlations  of the state $\rho_{W\bar{W}}$ is violated i.e., 
\be
D_{AB}+D_{AC}=2 D_{AB}\approx 2 \times 0.386 > D_{A:BC} \approx 0.45
\ee
In other words, the three qubit pure state $\rho_{W\bar{W}}$ is polygamous.
Thus, despite belonging to the same SLOCC class ${\cal D}_{1,1,1}$, the states $\vert {\rm GHZ}\rangle$ and $\vert {\rm{W{\bar W}}}\rangle$ exhibit contrasting behaviour corresponding to the shareability of correlations amongst subsystems. 
It is pertinent to point out yet another distinguishing feature of these states~\cite{WL1,WL2}:  the three qubit W-superposition state $\vert {\rm{W{\bar W}}}\rangle$  possesses reducible correlations~\cite{usrmaj, usa2} while the correlations in the GHZ state are not attributable to its subsystems.  The monogamous nature of GHZ state and the polygamous nature of W-superposition state brought out here (in terms of the correlation measure RRQD) thus supports the assertion that the states belonging to the same SLOCC class can exhibit quite dissimilar features --  especially  with respect to the reducibility and shareability of correlations.  

It is worth pointing out at this juncture that a possibility of using mono/polygamous nature with respect to quantum discord as a witness to distinguish states belonging to different SLOCC classes is discussed recently in \cite{prabhu}. Here, in this work, polygamy with respect to RRQD is identified to be a necessary criterion for $3$ qubit pure symmetric states belonging to the SLOCC class ${\cal{D}}_{2,1}$. This provides a clear signature for distinguishing states belonging to this class using RRQD as a witness.  

\section{Generalized GHZ and W states: Examination of monogamous nature} 

After examining the two SLOCC classes of $3$ qubit pure symmetric states for mono/polygamous nature, it is of interest to check how non-symmetric states, generalizing the family of symmetric three qubit states of the distinct SLOCC classes, behave in the context of  monogamy with respect to RRQD. The non-symmetric states under consideration here are the generalized GHZ states of the form 
\be
\label{genghz1}
\vert {\rm{GHZ}}\rangle_{\rm gen}=a \vert 000 \rangle+ b \vert 111 \rangle,\ \ a=|a| e^{i\alpha}, \ b=|b| e^{i\beta}; \ \ |a|^2+|b|^2=1
\ee 
and the generalized W states of the form 
\begin{eqnarray}
\label{genw1}
\vert {\rm{W}}\rangle_{\rm gen}&=&a \vert 100 \rangle+ b \vert 010 \rangle+ c \vert 001 \rangle, \\ 
& & a=|a| e^{i\alpha}, \ b=|b| e^{i\beta}  \ \ c=|c| e^{i\gamma};\ \  |a|^2+|b|^2+|c|^2=1. \nonumber 
\end{eqnarray}
While $\vert {\rm{GHZ}}\rangle_{\rm gen}$ are related to GHZ states belonging to the SLOCC class of $3$ distinct spinors, $\vert {\rm{W}}\rangle_{\rm gen}$ are related to W states belonging to the SLOCC class of $2$ distinct spinors. 

\subsection{Generalized GHZ states} 

Inspite of the inherent non-symmetry in the state $\vert {\rm{GHZ}}\rangle_{\rm gen}$ (owing to the fact that $a\neq b$ in  Eq. (\ref{genghz1})), one can find that 
\be
\rho_{AB}=\rho_{AC}=\rho_{BC}=\ba{cccc} |a|^2 & 0 & 0 & 0 \\ 0 & 0 & 0 & 0  \\ 0 & 0 & 0 & 0 \\ 0 & 0 & 0 & |b|^2 \ea
\ee
and 
\be
\rho_{A}=\rho_{B}=\rho_{C}=\ba{cc} |a|^2 & 0 \\ 0 & |b|^2 \ea.
\ee
Thus, it can be easily seen that $D_{AB}=D_{AC}=D_{BC}=0$ quite similar to that in the case of $\vert {\rm{GHZ}}\rangle$. 

The completely decohered counterpart of $\rho_{{\rm GHZgen}}=\vert {\rm GHZ} \rangle_{\rm gen} \langle {\rm GHZ} \vert$ is given by 
\be
\rho^d_{{\rm GHZgen}}=\mbox{diag} \left(|a|^2,\,0,\,0,\,0,\,0,\,0,\,0,\,|b|^2 \right)
\ee
and hence 
\begin{eqnarray}
\label{ghzg}
D_{A:BC}&=&0-\left(|a|^2 \log |a|^2+|b|^2 \log |b|^2\right) \nonumber \\ 
&=&-\left(|a|^2 \log |a|^2+(1-|a|^2) \log (1-|a|^2) \right), \ \  0\leq |a| \leq 1.
\end{eqnarray}
Clearly, $Q_{{\rm II}_{\rm gen}}=D_{AB}+D_{AC}-D_{A:BC}=\left(|a|^2 \log |a|^2+(1-|a|^2) \log (1-|a|^2) \right)>0$ implying the monogamous nature of the generalized GHZ states with respect to RRQD. This behaviour is in accordance with their symmetric counterpart $\vert {\rm GHZ} \rangle$ belonging to the class ${\cal D}_{1,1,1}$, discussed in Sec.IV (B). Thus monogamy with respect to RRQD is a necessary condition for states to be of the generalized GHZ type.  

\subsection{Generalized W states}

The generalized W state, unlike $\vert {\rm GHZ} \rangle_{\rm gen}$, possesses unequal subsystem density matrices.  Thus, both $D_{AB}$, $D_{AC}$ are to be evaluated. The non-zero eigenvalues of $\rho_{AB}$ being given by $|a|^2$, $|b|^2+|c|^2$ and that of its completely decohered counterpart $\rho^d_{AB}$ being given by $|a|^2$, $|b|^2$ and $|c|^2$, one gets, 
\begin{eqnarray}
D_{AB}&=&|a|^2 \log |a|^2+(|b|^2+|c|^2) \log (|b|^2+|c|^2)-\left(|a|^2 \log |a|^2+|b|^2 \log |b|^2+|c|^2 \log |c|^2 \right) \nonumber \\
&=& (|b|^2+|c|^2) \log (|b|^2+|c|^2)-\left(|b|^2 \log |b|^2+|c|^2 \log |c|^2\right). 
\end{eqnarray}
Similarly, the non-zero eigenvalues of  $\rho_{AC}$ are given by  $|b|^2$, $|a|^2+|c|^2$ and that of $\rho^d_{AC}$ are given by $|a|^2$, $|b|^2$  and  $|c|^2$. Thus, 
\begin{eqnarray}
D_{AC}&=&|b|^2 \log |b|^2+(|a|^2+|c|^2) \log (|a|^2+|c|^2)-\left(|a|^2 \log |a|^2+|b|^2 \log |b|^2+|c|^2 \log |c|^2\right) \nonumber \\
&=& (|a|^2+|c|^2) \log (|a|^2+|c|^2)-\left(|a|^2 \log |a|^2+|c|^2 \log |c|^2\right). 
\end{eqnarray}
On evaluating the completely decohered counterpart of $\rho_{ABC}$, in the eigenbasis of the subsystems $A$, $BC$ (following Sec. IV (B)), one can see that its diagonal elements (eigenvalues) are dependent on the angles $\alpha$, $\gamma$ also. On obtaining the explicit expression for $D_{A:BC}$, and on eliminating $|c|$ using $|c|=\sqrt{1-|a|^2-|b|^2}$, $Q_{{\rm II}_{\rm gen}}$ is arrived at as a function of $|a|$, $|b|$, $\alpha-\gamma$. 
A three dimensional plot of $Q_{{\rm II}_{\rm gen}}$ as a function of $|a|$, $|b|$, for the case of $\alpha=\gamma$ is given in Fig. 2. 
\begin{figure}[h]
\includegraphics*[width=2.5in,keepaspectratio]{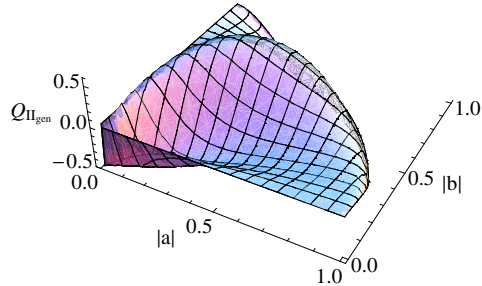}
\caption{The $3$-dimensional plot of $Q_{{\rm II}_{\rm gen}}=D_{AB}+D_{AC}-D_{ABC}$ for $3$ qubit generalized W states. Depending on the parameters $|a|$, $|b|$, the state is either monogamous or polygamous.} 
\end{figure}
It can be seen through the figure that generalized W states exhibit both mono and polygamous nature unlike that of their symmetric counterpart $\vert W\rangle$ belonging to the class ${\cal D}_{2,1}$. While the monogamous nature obeyed by GHZ state is retained by its non-symmetric, generalized counterpart $\vert {\rm GHZ} \rangle_{\rm gen}$, it is not so in the case of W state and its generalized counterparts. In fact, the monogamous nature of the GHZ states and their generalized, non-symmetric counterparts, allows for an identification of this class of states using monogamy with respect to RRQD as a witness.

\section{Summary and concluding remarks} 
This is a contribution towards an important issue whether monogamy of correlations holds for correlation measures other than entanglement by examining RRQD as another measure. The present work gives an overall picture of RRQD in comparison with quantum entanglement and also relates to SLOCC classification as well. 

The results obtained are displayed
succinctly in Table I.
\begin{table}[ht] 
\caption{}
\begin{tabular}{|c|c|c|c|c|c|c|}
\hline
SLOCC & State & ${\cal D}_{2,1}$ & ${\cal D}_{1,1,1}$ & Monogamous & Polygamous & Entanglement type \\ 
\hline
Yes & $\vert \psi\rangle$ (Eq.\ref{sim}) &  Yes & & & Yes  & No 3-way only 2-way \\
\hline 
\cline{2-7}
 Equivalent & $\vert {\rm GHZ}\rangle$ (Eq.(\ref{ghz}))  &     & Yes & Yes  & & No 2-way, only 3-way \\ 
  \cline{2-7}
 & $\vert \rm{ W\bar{W}} \rangle$ (Eq.(\ref{ww}))& & Yes & & Yes &   Both 3-way and 2-way \\
 \hline 
 \cline{2-7}
Not & $\vert {\rm GHZ}\rangle_{\rm gen}$ (Eq.\ref{genghz1}) &  & Yes  & Yes & &  No 2-way, only 3-way \\ 
\cline{2-7}
Equivalent & $\vert {\rm W}\rangle_{\rm gen}$ (Eq.\ref{genw1}) & Yes  & & Yes (parameter & Yes (parameter &   No 3-way, only 2-way \\ 
& & & &  dependent) & dependent) & \\ 
\hline
\end{tabular}
\end{table} 

Pure three qubit states are the focus here and they are described in a unified way by employing their Majorana representation.  It is found that the states belonging to the SLOCC class ${\cal D}_{2,1}$ are polygamous while those belonging to the class 
${\cal D}_{1,1,1}$ exhibit contrasting mono, polygamous behaviour with respect to RRQD. 
The nonsymmetric  GHZ states, the so-called generalized GHZ states, too are found to be monogamous. The generalized W states exhibit parameter dependent mono-, polygamous behaviour unlike their symmetric counterparts, the W states which are polygamous as they belong to ${\cal D}_{2,1}$ of two distinct spinors. 

An important feature that arises due to the polygamous behaviour of the states belonging to ${\cal D}_{2,1}$  and the monogamous nature of generalized GHZ states is the use of RRQD as a witness to identify states belonging to these classes. RRQD being an operationally convenient measure of quantum correlations, not requiring any optimization procedures for its evaluation, it is preferable to use it for this purpose.

\end{document}